\documentclass[11pt]{article}
\pdfoutput=1
\usepackage{jheppub}
\usepackage{rotating}
\usepackage{enumitem}
\usepackage{wrapfig}
\usepackage[left=5cm,right=0cm,top=5cm,bottom=1cm,footskip=1cm]{geometry}

\usepackage{amsfonts}
\usepackage{amsmath}
\usepackage{slashed}
\usepackage{amssymb}
\usepackage{latexsym}
\usepackage{multirow}
\usepackage{hyperref}
\usepackage{enumitem}
\usepackage{mathtools}
\hypersetup{colorlinks=true,citecolor=red,linkcolor=magenta,urlcolor=blue}
\usepackage{relsize}
\usepackage{booktabs}
\usepackage{subfigure}
\usepackage{cleveref}
\usepackage{fancyhdr}
\usepackage{float}
\usepackage{graphicx}
\usepackage{microtype}
\usepackage{tabularx}
\usepackage[dvipsnames]{xcolor}
\usepackage{orcidlink}
\usepackage{wrapfig}
\usepackage{siunitx,adjustbox,bm}
\usepackage[bottom]{footmisc}
\usepackage{placeins,bbold}
\renewcommand{\d}{\text{d}}
\renewcommand{\vec}[1]{{\bf{#1}}}

\title{Polyakov Loops Tame Phase Transitions}
\author[a]{Lisa Biermann\orcidlink{0000-0003-1469-6400},} 
\author[b]{Joydeep Chakrabortty\orcidlink{0000-0001-8709-916X},}
\author[c]{Christoph Englert\orcidlink{0000-0003-2201-0667}} 
\affiliation[a]{PSI Center for Neutron and Muon Sciences, 5232 Villigen PSI, Switzerland}
\affiliation[b]{Indian Institute of Technology Kanpur, Kalyanpur, Kanpur 208016, Uttar Pradesh, India}
\affiliation[c]{Department of Physics \& Astronomy, University of Manchester, Manchester M13 9PL, United Kingdom}
\emailAdd{lisa.biermann@psi.ch}
\emailAdd{joydeep@iitk.ac.in}
\emailAdd{christoph.englert@manchester.ac.uk}
\abstract{We estimate the impact of Polyakov loop (PL) contributions on electroweak phase transitions (PTs).  We show that the PL, which is unavoidable in thermal gauge field theory, tends to tame thermal contributions, thereby softening electroweak PTs and affecting bubble dynamics, nucleation, and the related gravitational-wave spectrum. Including this 
contribution in perturbative approaches results in a thermal effective potential that disfavours first-order PTs over either second-order PTs or smooth cross-overs. This feature is universal for both fermionic and bosonic contributions to the effective potential.}
\begin{document}
\allowdisplaybreaks
\flushbottom
\renewcommand{\ln}{\log}
\maketitle
\allowdisplaybreaks
\section{Introduction}
\label{sec:intro}
Electroweak phase transitions (PTs) and related dynamics are key drivers for the phenomenology of our early universe. They are particularly relevant when considered in conjunction with the known shortcomings of the Standard Model (SM) of Particle Physics, in which first-order electroweak PTs are critical for, e.g., electroweak baryogenesis to fulfil Sakharov's criteria~\cite{Sakharov:1967dj}. The available tools for analysing the electroweak thermal evolution often rely on perturbatively improved effective potential calculations~\cite{Wainwright:2011kj,Basler:2018cwe,Athron:2020sbe,Basler:2020nrq,Ekstedt:2022bff,DRansitions,Basler:2024aaf,Athron:2024xrh,Brdar:2025hxw,Costa:2025pew,BLOOP,Matuszak:2026xsz,Wang:2026jjn}. Computations using lattice approaches are considered in parallel \cite{Kajantie:1995kf,Cline:1996cr,Laine:1998jb,Ratti:2005jh,Kainulainen:2019kyp,Gould:2019qek,Gould:2022ran,Niemi:2024axp}, and are also compared to higher-order improved perturbative methods in e.g.~\cite{Gould:2021dzl,Ekstedt:2022zro,Gould:2023ovu,Ekstedt:2024etx}. By computing the `bounce action'~\cite{Coleman:1977py,Callan:1977pt,Linde:1977mm,Affleck:1980ac,Linde:1981zj,Konstandin:2006nd,Ekstedt:2021kyx,Gould:2021ccf,Lofgren:2021ogg,Hirvonen:2021zej,Kierkla:2025qyz}, statistical ensembles are then constructed to identify the correct vacuum configuration and gauge the transition rate. In the context of beyond-the-SM (BSM) scenarios, such an analysis forms a critical part of baryon-number generation that is additionally informed by bubble dynamics, CP violation, and sphaleron conversion and washout (for reviews see~e.g.~\cite{Morrissey:2012db,vandeVis:2025efm}).

Currently available tools~\cite{Wainwright:2011kj,Basler:2018cwe,Athron:2020sbe,Basler:2020nrq,Ekstedt:2022bff,DRansitions,Basler:2024aaf,Athron:2024xrh,Brdar:2025hxw,Costa:2025pew,BLOOP,Matuszak:2026xsz,Wang:2026jjn} allow one to study the PT in terms of a classical scalar (Higgs) background field, relying on the effective potential and analysing the respective local and global minima. These approaches, primarily based on perturbative methods in diagrammatic and functional techniques, do not, in general, capture all relevant effects. A pioneering example of the latter is the so-called Polyakov loop (PL)~\cite{Polyakov:1975rs,Polyakov:1978vu,Susskind:1979up,Weiss:1980rj} (a Wilson loop in thermal field theory).  In Ref.~\cite{Weiss:1980rj}, the PL is identified as an order parameter of (de)confinement in QCD, and it is demonstrated that the vanishing trace of the PL implies that the system is in the confined phase; otherwise, it is in the deconfined phase. 
Loop-improved effective actions for the PL in pure gauge theories have been discussed in \cite{KorthalsAltes:1993ca,Dumitru:2013xna,Nishimura:2018wla} using constraint potential techniques~\cite{Fukuda:1974ey,ORaifeartaigh:1986axd}.
Thermal Wilson lines, more broadly, characterise how coloured excitations propagate in the quark-gluon plasma (see e.g.~\cite{Gross:1980br,Pisarski:2000eq,Pisarski:2002ji}). It has been noted that the thermal partition function for a spin system requires a non-zero imaginary chemical potential that ensures the removal of unphysical degrees of freedom \cite{Popov:1988fdi}. This imaginary chemical potential $(\mu_I)$ is related to the gauge charges and is quantised \cite{Popov:1988fdi,Veits-Oppermann-Binderberger-Stein}. The non-trivial presence of the PL is mandatory for consistent quantisation in thermal gauge field theory and for compatibility with periodic boundary conditions for fields in the thermal partition function \cite{Weiss:1980rj}. Furthermore, given the results of \cite{Chakrabortty:2024wto, Bandyopadhyay:2026nrv}, the PL in thermal gauge field theory (in the absence of dynamical matter fields) replicates the behaviour of the imaginary chemical potential $\beta \mu_I$ of spin chains, see Refs.~\cite{Weiss:1980rj,Popov:1988fdi,Veits-Oppermann-Binderberger-Stein,Svetitsky:1982gs, Chakrabortty-Ghosh-Kundu}. 

In this work, we consider modifications to the perturbative effective potential by including PL contributions and develop a framework for their inclusion in publicly available codes. Considering a toy scenario, we trace the relevance of the PL to phenomenological implications such as the strength of the first-order PT and its interplay with bubble nucleation rates, as well as expected gravitational-wave spectra. This paper is organised as follows: In Sec.~\ref{sec:details}, we discuss how the effective potential is modified by the inclusion of the PL. We develop a practical way to include these effects by modifying the well-known characteristic functions of the $T>0$ domain. Subsequently, in Sec.~\ref{sec:model}, we perform a case study inspired by existing analyses in the literature to highlight the phenomenological relevance of the PL in a motivated toy scenario. Implications for bubble nucleation and the resulting gravitational-wave phenomenology are discussed. We conclude in Sec.~\ref{sec:conc}.

\section{Polyakov Loops and their Imprint on the Effective Potential}
\label{sec:details}
We start with a brief review of the relevance of Polyakov loops (PLs)~\cite{Polyakov:1975rs,Polyakov:1978vu} in thermal field theory and how they lead to corrections to the effective potential used to analyse PT dynamics. At zero temperature, Quantum Field Theory provides a framework for computing transition probabilities. 
In the early Universe at non-zero temperature, however, the background state is a thermal bath, and the dynamics are described within thermal field theory. 
The transition to this setting is implemented by euclideanising Minkowski time, $t\to i\beta$, with inverse temperature $\beta=1/T$~\cite{Matsubara:1955ws,Niemi:1983nf,Landsman:1986uw}. Green's functions reflect this transition through the Kubo-Martin-Schwinger~\cite{Kubo:1957mj,Martin:1959jp} relation. In the imaginary time formalism that can be used to tackle practical calculations~\cite{Matsubara:1955ws}, this imposes (anti)periodicity conditions on the fields of the theory. Put differently, thermal field theory lives on $S^1\times {\mathbb{R}}^3$ where the compactification radius is determined by $\beta$. This space is topologically non-trivial as $\pi_1(S^1\times {\mathbb{R}}^3)=\mathbb{Z}$. 
Just as for $T=0$, non-trivial topology can have non-trivial non-perturbative implications. In thermal gauge theory, these are encoded in the holonomy around the Euclidean time circle, the PL matrix, i.e., a thermal Wilson~loop
\begin{equation}
  {\bf{L}} (\vec{x}) =  \mathbf{T} \left(\exp\left\{-\int_{0}^{\beta} d\tau A_0 (\vec{x},\tau)\right\}\right )\,,
\end{equation}
where $\mathbf{T}$ defines Euclidean time path-ordering and $A_0$ is the temporal component of the Euclidean gauge field. The PL is the thermal equivalent of a Wilson line that is wrapped around Euclidean time on the circle $S^1$. A gauge-invariant diagnostic is the traced (normalised) PL
\begin{equation}
  \Omega (\vec{x}) = \frac{1}{N}\,\text{Tr}\,{\bf{L}}\,,
\end{equation}
(the trace is evaluated on the gauge indices), which serves as an order parameter for (de)confinement in pure $SU(N)$ gauge theory~\cite{Weiss:1980rj}. Perturbative calculations typically assume a specific gauge choice, but $A_0$ cannot be removed globally, because it is determined by holonomy. 

We consider a scalar ($\varphi$) quantum field theory with a background $SU(N)$ gauge theory.  Here, we employ the Heat-Kernel (HK) method, defined in Euclidean space, to compute the one-loop effective potential after integrating out the quantum fluctuations associated with the scalar field.\footnote{It is worth mentioning that, as the gauge fields are in the background, there are no gauge quantum fluctuations, and thus they do not contribute to the one-loop effective potential directly.} The one-loop effective Lagrangian in $(d+1)$-dimensional Euclidean spacetime\footnote{We work with an all-negative metric.} ${\mathbb{R}}^{d+1}$ is given by~\cite{Chakrabortty:2024wto, Bandyopadhyay:2026nrv}
\begin{equation}\label{eq:Leff-1L}
    \mathcal{L}^{1L} = c_s\; \text{Tr}\; \int_0^\infty \frac{\d t}{t} \int \frac{\d^{d+1}p}{(2\pi)^{d+1}}\; e^{-m^2t}\; 
    \langle x| e^{-(D^2+U)t} | p \rangle \langle x|p\rangle\,,
\end{equation}
where $\langle x|p\rangle = \exp\{ix_\mu p^\mu\}$, $c_s=1/2~(1)$ for real (complex) scalars and $-1$ for Dirac fermions. $D_\mu=\partial_\mu+A_\mu$ denotes the covariant derivative with $A_\mu$\footnote{The respective gauge coupling is absorbed in the gauge field in this convention.} as the Euclidean gauge field in ${\mathbb{R}}^{d+1}$, and $D^2 = D_\mu D^\mu$. We define $U={\delta ^2V^{\text{tree}}}/{\delta \varphi^2}$, where $V^{\text{tree}}$ is the tree-level potential of $\varphi$. From this we can obtain thermal field theory on $ S^1 \times {\mathbb{R}}^d $ with bosonic and fermionic fields that satisfy periodic and anti-periodic boundary conditions on $S^1$, respectively~\cite{Megias:2003ui, Moral-Gamez:2011wcb, Chakrabortty:2024wto}, leading to
\begin{equation}
    \mathcal{L}^{1L} = c_s\; \frac{1}{\beta}  \int_0^\infty \frac{\d t}{t} \;\text{Tr}\; \sum_{p_0} \int_0^\infty \frac{\d^{d}p}{(2\pi)^{d}}\; e^{-m^2t}\;  e^{-((D_i+p_i)^2-Q^2+U)t}\,,
\end{equation}
where $\langle \vec{p}, p^0| \vec{k},k_0 \rangle = (2\pi)^d\, \beta\, \delta(\vec{p}-\vec{k}) \delta_{p_0,k_0}$, and $p_0={2\pi n}/{\beta}$ with $n\in \mathbb{Z}$ is the quantised momentum along $S^1$ (referred to as the Matsubara modes~\cite{Matsubara:1955ws}). We further define $Q^2=|D_0+ip_0|^2=|\partial_0+A_0+ip_0|^2$ with $A_0$\footnote{By convention, we absorb the coupling constant into the gauge field, $A_0=g A_{0,j} T^j$, where $g$ is the gauge coupling, $T_j$ are the generators in the fundamental representation, where $j=1,\cdots, (N^2-1)$ for an $SU(N)$ gauge theory.} as the Euclidean (imaginary) temporal gauge field and $A_0=\text{const.} \neq 0$ being a consistent choice~\cite{Weiss:1980rj}.
With this, the PL for an $SU(N)$ gauge theory reads
\begin{equation}
  {\bf{L}}=  \exp\{-\beta A_0\} \Rightarrow  \Omega  =  \frac{1}{N} \, \text{Tr} \exp\{-\beta A_0\}\,,
\end{equation}
which leads to
\begin{equation}
\label{eq:qdef}
Q^2=\Big|i \frac{2n\pi}{\beta} +A_0\Big|^2=\Big|i \frac{2n\pi}{\beta} -\frac{\ln\;  {\bf{L}}}{\beta}\Big|^2= \Big(\frac{2\pi}{\beta}\Big)^2\left[n+i\,\frac{A_0\beta}{2\pi}\right]^2\,.
\end{equation}
Now, within this thermal field theory framework with gauge fields in the background, the gauge-invariant and gauge-parameter-independent one-loop effective potential, including the PL contribution, can be computed using Eq.~\eqref{eq:Leff-1L} as~\cite{Chakrabortty:2024wto, Balui:2025yvd, Bandyopadhyay:2026nrv} 
\begin{align}
\label{eq:matsu-bo}
    -V_{\text{eff}}^{\text{1L},S} (m,\beta,\tilde n) &=  \frac{c_s }{\beta}\; \text{Tr}\; \;\int_0^\infty \frac{\d t}{t}\;  \sum_{p_0} \frac{e^{-m^2t}}{(4\pi t)^{\frac{d}{2}} }\; e^{Q^2t}\nonumber \\
    &= \frac{c_s}{\beta } \sum_n \frac{\mu^{4-d}}{(4\pi)^{d/2}}\; \Gamma(-d/2)\;\text{Tr}\; \left[ m^2+Q^2 \right]^{d/2}\nonumber\\
        &= \frac{c_s}{\beta } \sum_n \frac{\mu^{4-d}}{(4\pi)^{d/2}}\; \Gamma(-d/2)\; \left[ m^2+\langle Q^2 \rangle \right]^{d/2}\nonumber\\
    &= \frac{c_s}{\beta } \sum_n \frac{\mu^{4-d}}{(4\pi)^{d/2}} \left[ m^2+\left(\frac{2\pi}{\beta} \right)^2 (n+\tilde{n})^2\right]^{d/2} \Gamma(-d/2)\nonumber\\
  &=  \frac{1}{2} \left [\frac{m^4}{32\pi^2} \left ( \ln \left( \frac{ 4\pi\mu^2}{m^2}\right)+\frac{3}{2}-\gamma_E\right) + \frac{m^2}{ \pi^2 \beta^2} \sum_{n=1}^\infty \frac{\cos (2\pi n \tilde{n})}{n^2} \mathbf{K}_2(m\beta n)   \right ]\,,
\end{align}
where $\mathbf{K}_2$ is the modified Bessel function of the second kind, $m$ is the background field-dependent mass of the scalar field, $\mu$ is the renormalisation scale, and $\gamma_E$ is the Euler constant.
Furthermore, 
\begin{equation}
   \langle Q^2\rangle = \left(\frac{2\pi}{\beta} \right)^2 \left\langle\left[n+i\frac{ A_0 \beta}{2\pi}\right]^2 \right\rangle=
    \left(\frac{2\pi}{\beta}\right)^2\left[n+\tilde{n}\right]^2\, ,
\end{equation}

\begin{table}[!t]
\centering
\begin{tabular}{|l|c|c|c|}
\hline
$R$ & $2$ & $3$ & $4$ \\
\hline
& & & \\[-12pt]
$\hat{A}_{0,3}$ & $\pm \dfrac{i\pi}{\beta}$ & $\pm \dfrac{2i\pi}{3\beta}$ & $\pm \dfrac{2i\pi}{4\beta}$, $\pm  \dfrac{6i\pi}{4\beta}$ \\[8pt]
\hline
& & & \\[-12pt]
$\tilde{n}= \mp \dfrac{i\beta \hat{A}_{0,3}}{4\pi}$ & $ \mp \dfrac{1}{4}$ & $\mp \dfrac{1}{6}$ & $\mp \dfrac{1}{8}$, \;\;$\mp \dfrac{3}{8}$ \\[8pt]
\hline
\end{tabular}
\caption{Values of $\hat{A}_{0,3}$ for different representations ($R$) of $SU(2)$ in the confined phase, $\Omega  =0$. Note that the effective potential is independent of the \textit{sign} of $\tilde{n}$. \label{tab:PL-Rep}}
\end{table}

In practice, the connection of $\tilde n$ and $\Omega$, i.e., $ \langle Q^2\rangle$ is established via the eigenvalues of the untraced PL, which we can bring to the maximal torus of the group~\cite{Weiss:1980rj}. Assuming, for simplicity, that the underlying gauge symmetry is $SU(2)$, we can express a representative gauge field, in Polyakov gauge,  $A_0 = \hat A_{0,3} \,{\mathbb{I}}_3$ with the diagonal isospin generator ${\mathbb{I}}_3$ of the $SU(2)$ algebra in the considered representation \cite{Salcedo:1998sv}. For an $SU(2)$ gauge theory with spin $J$ representation ($R=2J+1$), the PL is
\begin{equation}\label{eq:omega}
\Omega= \frac{1}{2}\text{Tr} \left[ \text{diag} \left\{ \exp(- J \beta \hat{A}_{0,3}),\cdots, \exp( J \beta \hat{A}_{0,3}) \right\} \right]\,.
\end{equation}
The PL contribution to the Matsubara sum, i.e., $\tilde{n}$ in Eq.~\eqref{eq:matsu-bo}, for pure gauge theory, is related to the PL as 
\begin{equation}
\tilde{n} =\frac{i\beta}{4\pi}\hat{A}_{0,3}\,,
\end{equation} 
cf.~Eq.~\eqref{eq:qdef}.\footnote{The additional factor of $1/2$ appears due to the normalisation $N=2$ for $SU(2)$. In the case of a semi-simple gauge theory $ \otimes_i\mathcal{G}_i$, with different gauge groups $\mathcal{G}_i$, we will have PL contributions for each gauge group if the scalar and/or fermion fields contain all gauge charges. In that case, $\tilde{n}=\mathcal{F}(\tilde{n}_i,\cdots,\tilde{n}_n)$, and the explicit form of this function $\mathcal{F}$ is model-dependent.}
In the absence of dynamical matter fields, i.e., pure gauge theory, the PL is the order parameter of (de)confinement, and confinement is characterised by the vanishing trace of the PL, i.e., $\Omega  =0$.
This corresponds to the `Weiss minima', which leave the $\mathbb{Z}_2$ centre symmetry invariant.\footnote{In the context of $SU(N)$ gauge theory, it is $\mathbb{Z}_N$.} The corresponding solutions also guarantee the statistical removal of the unphysical (negative norm) degrees of freedom from the physical spectrum in the thermal gauge field theory. It is interesting to note that these solutions, see Tab.~\ref{tab:PL-Rep}, replicate the imaginary chemical potential solutions that are the result of the Popov-Fedotov trick for algebraic removal of unphysical degrees of freedom of spin chains \cite{Popov:1988fdi,Veits-Oppermann-Binderberger-Stein}. 

We highlight that the presence of a background gauge field enters the effective potential only through the PL via $\tilde{n}$. In the case of a scalar gauge theory, where the gauge fields, like the scalars, are dynamical, we must integrate out the gauge-field fluctuations over their respective backgrounds. This leads to an additional contribution to the effective potential that contains the PL effect only for the non-Abelian case.
This result has been generalised to include the contributions from dynamical gauge fields for Abelian scalar QED, and the PL-corrected one-loop effective potential has been computed in~\cite{Bandyopadhyay:2026nrv}. We base our work on this result.

As in the conventional diagrammatic approach that sums all one-particle-irreducible vertex functions, Eq.~\eqref{eq:matsu-bo} expresses the effective potential as a sum over Matsubara modes\footnote{It is worth highlighting that infrared (IR) problems associated with the zero modes present for bosons are ameliorated when $\tilde n\neq 0$, as the PL contribution shifts all modes by a real-valued, non-integer number $\tilde{n}$~\cite{Chakrabortty:2024wto}, i.e. the PL gives rise to a non-perturbative contribution to the Debye mass (see also~\cite{Arnold:1995bh,Laine:1999hh} for related discussions in QCD).} and recovers the Coleman-Weinberg potential~\cite{Coleman:1973jx} for $T=0$. For $\tilde n=0$, this sum can be expressed in a closed analytical form related to the dimensionless characteristic function 
\begin{equation}
\label{eq:functionsB0}
J_{B}(m,\beta, \tilde{n}=0) = \int_0^\infty \d x\, x^2 \log\left(1 - \exp\left\{-\sqrt{x^2 + m^2\beta^2}\right\}\right)\,,
\end{equation}
see~\cite{Quiros:1994dr,Quiros:1999jp}. This forms the starting point of perturbative numerical investigations using publicly available tools~\cite{Wainwright:2011kj,Basler:2018cwe,Athron:2020sbe,Basler:2020nrq,Basler:2024aaf,Athron:2024xrh,Brdar:2025hxw,Costa:2025pew,Matuszak:2026xsz,Wang:2026jjn}.
Including PL effects, we define (see also~\cite{Chakrabortty:2024wto})
\begin{equation}
\label{eq:functionsb}
J_B(m,\beta, \tilde{n}) = -\pi^2\beta^4\, S^{\tilde{n}}_\Omega(m,\beta)\,, \\
\end{equation}
with
\begin{equation}
\label{eq:stilde}
S^{\tilde{n}}_\Omega(m,\beta)
= \frac{m^2}{\pi^2\beta^2}\sum_{n=1}^{\infty}\frac{1}{n^2}\cos(2\pi n\tilde{n})\,{\mathbf{K}}_2(mn\beta)\,,
\end{equation}
which yields Eq.~\eqref{eq:functionsB0} for $\tilde n =0$. 
As numerical tools are built around Eq.~\eqref{eq:functionsB0} in a modular fashion (we discuss fermions further below), comparing Eqs.~\eqref{eq:functionsB0} and~\eqref{eq:functionsb} offers an economical strategy for incorporating PL corrections into existing frameworks.

In our analysis, we also consider dynamical matter fields. Thus, it is not necessarily the case that $\Omega=0$ satisfies the dual minimisation of the effective potential $V_{\text{eff}}^{\text{1L}}(\varphi, \tilde{n})$ with respect to $\varphi$ and $\hat{A}_{0,3}$. Therefore, for a given theory, the choice of $\tilde{n}$ in the effective potential should be consistent with $\delta V_{\text{eff}}^{\text{1L}}/\delta \varphi=\delta V_{\text{eff}}^{\text{1L}}/\delta \hat{A}_0^3=0$. In this work, our main objective is to display the impact of the PL on a potential PT. Thus, we include the impact of temporal gauge fields through the PL by assigning phenomenological values to $\tilde{n}$ instead of restricting to a confining gauge background. In other words, we will treat $\tilde n$ as a free constant\footnote{We choose it to be a (dimensionless) real number $<1 $, motivated by the solutions of $\Omega=0$ in Tab.~\ref{tab:PL-Rep}.} parameter that parametrises the plasma through this connection (see~\cite{Hidaka:2009ma,Pisarski:2000eq} for similar discussions in the context of the semi-quark-gluon plasma in QCD).

The effective potential after integrating out the chiral fermions in an $SU(N)$ background gauge theory is given by~\cite{Chakrabortty:2024wto}
\begin{align}
\label{eq:matsu-fermion}
    -V_{\text{eff}}^{\text{1L},F} &=   -\frac{d}{2} \bigg[\frac{m^4}{32\pi^2} \left ( \ln \left( \frac{ 4\pi\mu^2}{m^2}\right)+\frac{3}{2}-\gamma_E\right) \nonumber \\
  &\hspace{1.5cm} + \frac{m^2}{ 2\pi^2 \beta^2} \sum_{n=1}^\infty \frac{\cos (2\pi n \tilde{n})}{n^2} \mathbf{K}_2(2m\beta n)   - \frac{m^2}{ \pi^2 \beta^2} \sum_{n=1}^\infty \frac{\cos (2\pi n \tilde{n})}{n^2} \mathbf{K}_2(m\beta n) \bigg]\,,
\end{align}
where $d=3$. Similarly to the bosonic case, we can define an extended characteristic temperature-dependent function
\begin{equation}
\label{eq:functionsFn}
J_F(m,\beta, \tilde{n}) = \pi^2\beta^4\left[2\, S^{\tilde{n}}_\Omega(m,2\beta) - S^{\tilde{n}}_\Omega(m,\beta)\right]\,,
\end{equation}
with $S^{\tilde{n}}_\Omega(m,2\beta)$ given in Eq.~\eqref{eq:stilde}. Again, this reduces to the well-known form
\begin{equation}
\label{eq:functionsF0}
J_{F}(m,\beta, \tilde{n}=0) = \int_0^\infty \d x\, x^2 \log\left(1 + \exp\left\{-\sqrt{x^2 + m^2\beta^2}\right\}\right)\,,
\end{equation}
and a relatively straightforward replacement of the characteristic $J_F$ function in existing numerical tools is possible.

Further details on the $\tilde n \neq 0 $ functions are given in App.~\ref{sec:moreJ}, where we also present a closed integral form based on an imaginary chemical potential $\mu_I=\mu_I(\tilde n)$, which might prove helpful in future applications.

\section{Modified Critical Behaviour}
\label{sec:model}
\subsection{A Toy Model and its Effective Potential}
\label{sec:model1}
\newcommand{\gtwo}{g}
To examine the impact of the PL contribution transparently, we trace its relevance for an $SU(2)$ toy model\footnote{We emphasise that our intention is not to propose a realistic model for a first-order electroweak PT, but to define a suitable scenario to highlight the PL's role in electroweak PTs clearly with minimal technical complexity.} that nevertheless captures relevant bosonic features of the SM (for instance, a first-order PT for light Higgs masses). Concretely,
\begin{equation}
\label{eq:lag}
    -\mathcal{L} = m^2 |\Phi|^2 + \lambda|\Phi|^4 - (D^\mu \Phi)^\dagger D_\mu\Phi + \frac14 {{W}}^{a\,\mu\nu}{{W}}^a_{\mu\nu}\,,
\end{equation}
with the Higgs field in the fundamental representation
\begin{equation}
\Phi 
=\frac{1}{\sqrt{2}}
\begin{pmatrix} \rho + i \eta\\ \zeta +  \omega +i\psi\end{pmatrix}\,.
\end{equation}
The covariant derivative is given by
\begin{equation}
D_\mu = \partial_\mu -i {\gtwo} {W}^a_\mu\,\frac{{\sigma^a}}{2}\,,
\end{equation}
where the three Pauli matrices are denoted by $\sigma^a$ ($a =1,2,3$) as usual. This model, modulo rephasing, admits trivial $|\langle \Phi \rangle|  \equiv \omega/\sqrt{2}=0$ and/or non-trivial $\omega\neq 0$ vacuum configurations at finite temperature.\footnote{The vacuum is additionally characterised by $\langle \rho\rangle,\langle\eta\rangle,\langle\zeta\rangle,\langle\psi\rangle=0$.} At zero temperature, we can identify
\begin{equation}
\omega^2 \stackrel{T\to 0}\longrightarrow v^2 = -\frac{m^2}{\lambda}>0\,,
\end{equation}
analogous to the SM Higgs sector. Thus, this toy model provides a motivated candidate theory for investigating the impact of the PL contribution and its phenomenological consequences.

In the following, we investigate the finite-temperature dynamics employing the {\tt{BSMPT}} code~\cite{Basler:2018cwe,Basler:2020nrq,Basler:2024aaf}. This aligns with conventional choices in the literature, e.g. those recently used in Refs.~\cite{Braathen:2025svl,Biekotter:2025npc}. Concretely, we consider the one-loop effective potential\footnote{In Refs.~\cite{Balui:2025kat, Balui:2025yvd, Balui:2026ghs}, it has been shown that a gauge-invariant (and gauge-fixing parameter-independent) realisation of the one-loop effective potential is identical to functional methods using the Landau gauge.} 
\begin{equation}
\label{eq:potential}
    V_\mathrm{eff}^\mathrm{1L,\,toy} \equiv V^\mathrm{tree} + V^\mathrm{CW} + V^\mathrm{CT} + V^\beta + V^\mathrm{daisy}\,,
\end{equation}
where the tree-level potential is given by Eq.~\eqref{eq:lag}, 
\begin{equation}
    V^\mathrm{tree} = m^2 |\Phi|^2 + \lambda |\Phi|^4\,.
\end{equation}
The zero-temperature Coleman-Weinberg potential~\cite{Coleman:1973jx} reads 
\begin{equation}
    V^\mathrm{CW} = \frac{1}{64 \pi^2} \sum_{X=S,G} S_X\,  m_X^4 \log\left(\frac{m_X^2}{\mu^2} - k_X\right)\,,
\end{equation}
with $S_X  = \{1,\, 3\}$ and $k_X = \{3/2,\,5/6\}$ for $X=\{S,G\}$. The masses $m_{X}$ are the eigenvalues of the scalar ($S$) and gauge-boson ($G$) mass matrices, i.e. the states that run in the loop. For the toy model of Eq.~\eqref{eq:lag}, these are given by
\begin{equation}
m^2_{\chi,\eta,\psi} =\lambda(\omega^2 -v^2)\,,
\quad 
m^2_{\zeta}
=3\lambda\omega^2 - \lambda v^2\,,
\quad 
m^2_{W^a} 
=\frac{\gtwo^2}{4} \omega^2\,,
\end{equation}
such that $M_h^2=2\lambda v^2$ for the physical Higgs boson. The finite-temperature part of the potential, including the PL contribution, following the discussion in Sec.~\ref{sec:details} and Ref.~\cite{Bandyopadhyay:2026nrv}, can be written as
\begin{equation}
\label{eq:vt}
    V^\beta = \sum_{X=S,G}  
    \dfrac{S_X}{2 \pi^2\beta^4} \, J_{B}(m_X,\beta,\tilde n)\,.
\end{equation}
To reflect the PL contributions as detailed above, the $\tilde n$-modified finite-temperature functions have been implemented in the {\tt{BSMPT}} code.\footnote{Modified Bessel functions with generally complex arguments are evaluated by employing the {\tt C++} wrapper~\cite{complex_bessel} of the original routines of~\cite{slatec}.} A range of cross-checks have been performed to evaluate the finite-temperature $\tilde n$-corrected effective potential. In practice, we find that truncating the series at $n\simeq 600$ in Eq.~\eqref{eq:stilde} provides an accurate description of the PT phenomenology without compromising numerical efficiency. In particular, the `standard' $\tilde n=0$ results using the closed form of the $J$ functions are accurately reproduced. We therefore opt for the truncated series definition in our numerical simulation.

The daisy corrections, collected in $V^\mathrm{daisy}$, are derived from the one-loop thermal masses $\overline{m}_{X}^2$ 
\begin{equation}
    V^\mathrm{daisy}=-\frac{1}{12\pi\beta}\sum_{X=S,G} \left[(\overline{m}_{X}^2)^{3/2} - (m_{X}^2)^{3/2}\right]\,.
\end{equation}
Daisy corrections are critical in the PL-free case, where they improve the infrared behaviour of the zero Matsubara modes. Polyakov contributions, however, as mentioned in Sec.~\ref{sec:details}, act as additional regulators. Lastly, the counterterm potential $V^\mathrm{CT}$ is included to fix the vacuum expectation value (VEV) and the masses at zero temperature to their leading-order values.

\subsection[Phase Transition Dynamics for $\tilde{n} \neq 0$]{Phase Transition Dynamics for $\bf{\tilde{n} \neq 0}$}
\label{sec:ps}
As the Universe cools, it can undergo PTs, depending on the underlying theory we use to describe its dynamics. In particular, first-order electroweak PTs, which are shown to be realised within a variety of BSM models, allow for promising avenues to explain baryogenesis and open new observational prospects with the advent of space-based gravitational-wave (GW) observatories, cf.~\cite{Croon:2023zay,Athron:2023xlk,vandeVis:2025efm} for recent reviews and references therein. The strength of a first-order PT is conventionally estimated by the ratio
\begin{equation}
\label{eq:ptstrength}
\xi_c \equiv {\omega_c^\mathrm{true} - \omega_c^\mathrm{false} \over T_c}\,, 
\end{equation}
i.e., it is given by the difference in VEVs of the scalar fields between the true and false minimum at the critical temperature $\omega_c=\sqrt{2}\langle\Phi\rangle(T_c)$ (when the non-trivial second minimum that emerges during the cooling of the Universe and the high-temperature minimum are degenerate, and separated by a potential barrier), measured in units of the critical temperature $T_c$.\footnote{The gauge-dependence of this measure and its implications have given rise to alternative considerations~\cite{Patel:2011th,Ekstedt:2020abj,Ekstedt:2022zro}, see also~\cite{DiLuzio:2014bua,Espinosa:2015qea}. It is worth highlighting that the gauge dependencies are absent in the background Fermi gauges at all orders, as demonstrated in~\cite{Balui:2025yvd, Balui:2025kat, Balui:2026ghs}. We will not consider this further in the present work.} For a sufficiently strong first-order PT with $\xi_c \gtrsim 1$, sphaleron washout in the broken phase is considered to be reasonably suppressed inside the forming true-vacuum bubble to satisfy the conditions required for, e.g., baryogenesis in more realistic theories than the one considered in this work. The key to a first-order PT is the formation of a potential barrier separating the false from the true vacuum, with $\omega\neq 0$, and $\omega \stackrel{T\to 0}\longrightarrow v$. Tunnelling between the vacua proceeds via the so-called bounce action~\cite{Coleman:1977py,Callan:1977pt,Hammer:1978xu,Linde:1981zj}, which triggers the nucleation of expanding true vacuum bubbles, cf.~\cite{Devoto:2022qen} for a review. This creates the out-of-equilibrium condition required by Sakharov's criteria. 

In the following, we will focus primarily on the behaviour of $\xi_c$ as a function of $\tilde n$ to understand how the Polyakov contribution affects the PT dynamics. That said, we will also investigate the impact of $\tilde n$ on the dynamics at the nucleation (the formation of one bubble per Hubble volume) and percolation (the formation of a network of bubbles and the large-scale conversion of 34\% of the comoving volume to the true minimum~\cite{Shante01051971}) stages of the PT. At the latter stage, we will focus on the effect of $\tilde{n}$ on the sourced GWs. Comparing these stages with their $\tilde n=0$ counterparts is relevant because it provides insights into how $\tilde n$ can change the PT timeline and lead to direct phenomenological implications. We furthermore stress that our choice of $\tilde{n}$ is driven solely by phenomenological interests, to demonstrate how the inclusion of the PL contribution affects the order of the PT and related phenomena.

\begin{figure}[!t]
    \centering
    \includegraphics[width=.55\linewidth]{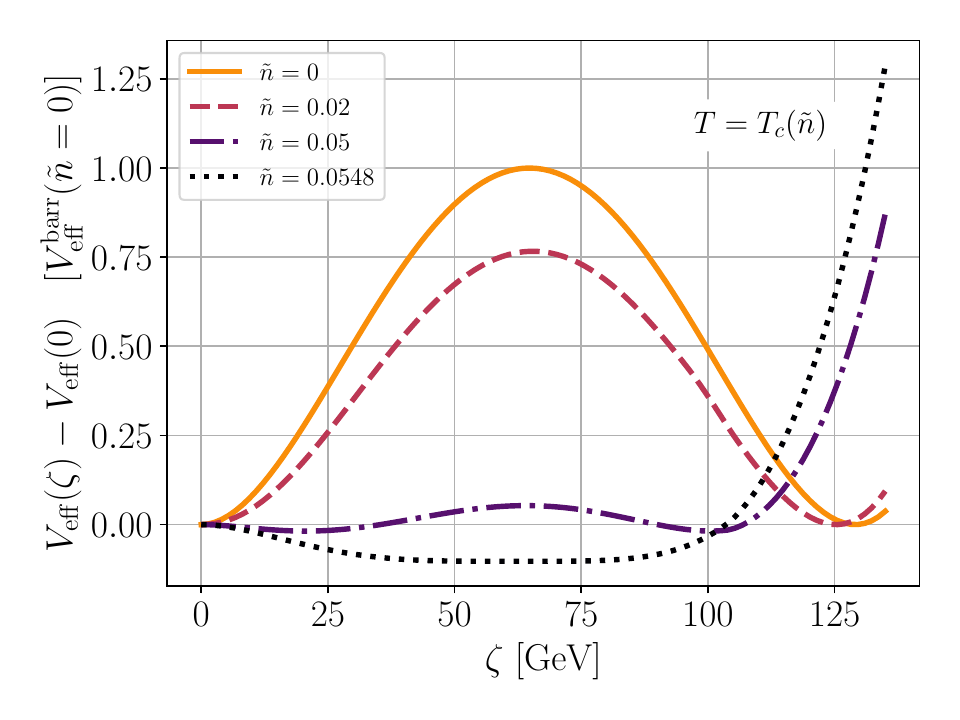}\hfill
    \includegraphics[width=.4\linewidth]{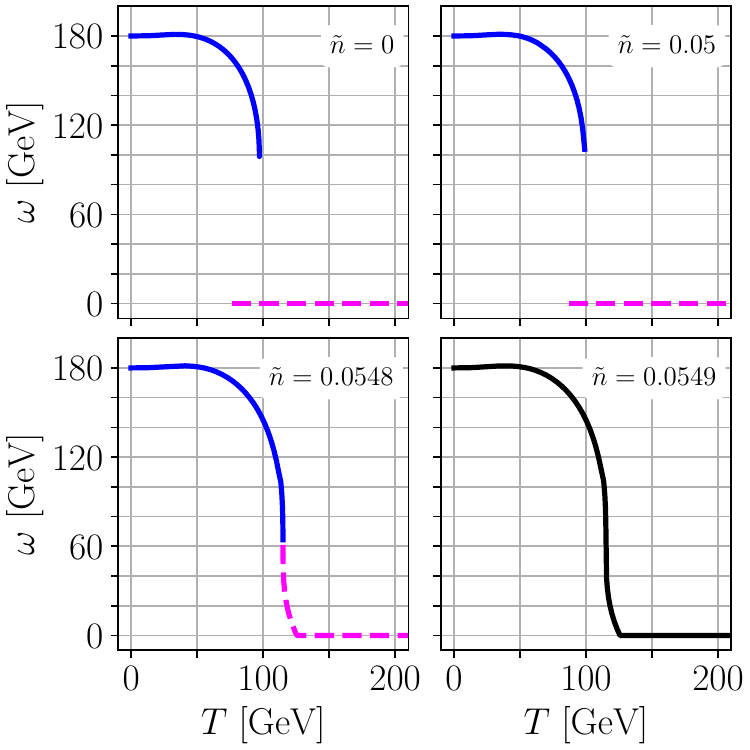}
    \caption{Impact of different values of $\tilde n$ on the BP of Eq.~\eqref{eq:bp}. Left: Potential contours at the critical temperatures normalised to the barrier height for $\tilde n = 0$. Right: Minima as a function of the temperature. Note that the discontinuity in the minima is reduced when $\tilde{n}$ is increased, which clearly reflects the smoothing of the barrier in the left panel.}
    \label{fig:bp_contours_phases}
\end{figure}

\subsubsection*{Phase Transitions with Bosonic Degrees of Freedom} 
Within the toy model detailed above, such a (strong) first-order PT can be observed for the $\tilde n =0$ potential towards smaller values of the scalar mass 
$M_h \in [\SI{30}{GeV},\,\SI{60}{GeV}]$ and larger values of the zero-temperature VEV $v \in [\SI{160}{GeV},\,\SI{200}{GeV}]$ for $g=\num{1.2}$. The motivation for this parameter choice is to maximise the parameter space compatible with a first-order PT. In doing so, we deliberately ignore other theoretical and experimental constraints of this toy model. This choice aligns with our aim to display and highlight the impact of the PL contribution on the parameter space compatible with a first-order PT.
We first consider the purely bosonic finite-temperature potential, then comment on the relevance of fermions below. We begin by discussing the impact of $\tilde n$ on the representative benchmark point (BP) with
\begin{align}
    M_h = \SI{46}{GeV},\quad v = \SI{180}{GeV},\quad g = \num{1.2}\,. \label{eq:bp}
\end{align}
In the left panel of Fig.~\ref{fig:bp_contours_phases}, we compare contours along the $\zeta$ direction of the effective potential at the critical temperature as a function of $\tilde n$. Starting with $\tilde n=0$, i.e. with the `standard' perturbative calculation, we observe a potential barrier that yields a first-order PT with $\xi_c\simeq \num{1.37}$. Increasing $\tilde n$ reduces the barrier height significantly compared to the $\tilde n=0$ case, and in addition leads to a shift of the low-temperature true minimum to lower field values and to an increase of the critical temperature. We find critical temperatures of $T_c=\{94.18,\,96.22,\,110.33,\,115.01\}\,\si{GeV}$ and resulting $\xi_c$ values of $\xi_c=\{1.37,\,1.30,\,0.72,\,0.07\}$ for $\tilde n = \{0,\,0.02,\,0.05,\,0.0548\}$. 
A value $\tilde n = \num{0.0548}$ is found to be the largest possible value that, for this BP, still yields a potential barrier.  This result is further illustrated by the right panel of Fig.~\ref{fig:bp_contours_phases}, where we show the temperature dependence of the found high- and low-temperature minima, as a dashed magenta and a solid blue line, respectively, for $\tilde n=\{0,\,0.05,\,0.0548,\,0.0549\}$. For $\tilde n = \num{0.0549}$, depicted by a solid black line, we find no barrier, but a continuous connection, compatible with~a second-order PT.\footnote{The distinction between second-order PT and smooth cross-over is beyond the scope of this work.} As also visible in Fig.~\ref{fig:bp_contours_phases}, $\tilde n$ changes the shape of the potential around the origin via odd monomials of the field-dependent mass, which consequently moves the high-temperature false minimum away from the origin to $\omega\neq0$. These monomials can be traced to the analytical properties at small $\tilde n$ that we detail in App.~\ref{sec:moreJ}.

From the in-depth analysis of the BP, we conclude that increasing $\tilde n$ not only softens the first-order PT, but can completely remove the barrier and turn the $\tilde n = 0$ expectation of a strong first-order PT into a second-order PT. The monotonic weakening of the PT with increasing $\tilde n$ is a universal feature and is expected to persist across models, rather than being specific to the toy model considered here. In this sense, although our analysis is carried out in a simplified setting, the qualitative behaviour we identify is expected to be broadly applicable.

\begin{figure}[!t]
\centering
\includegraphics[width=\textwidth]{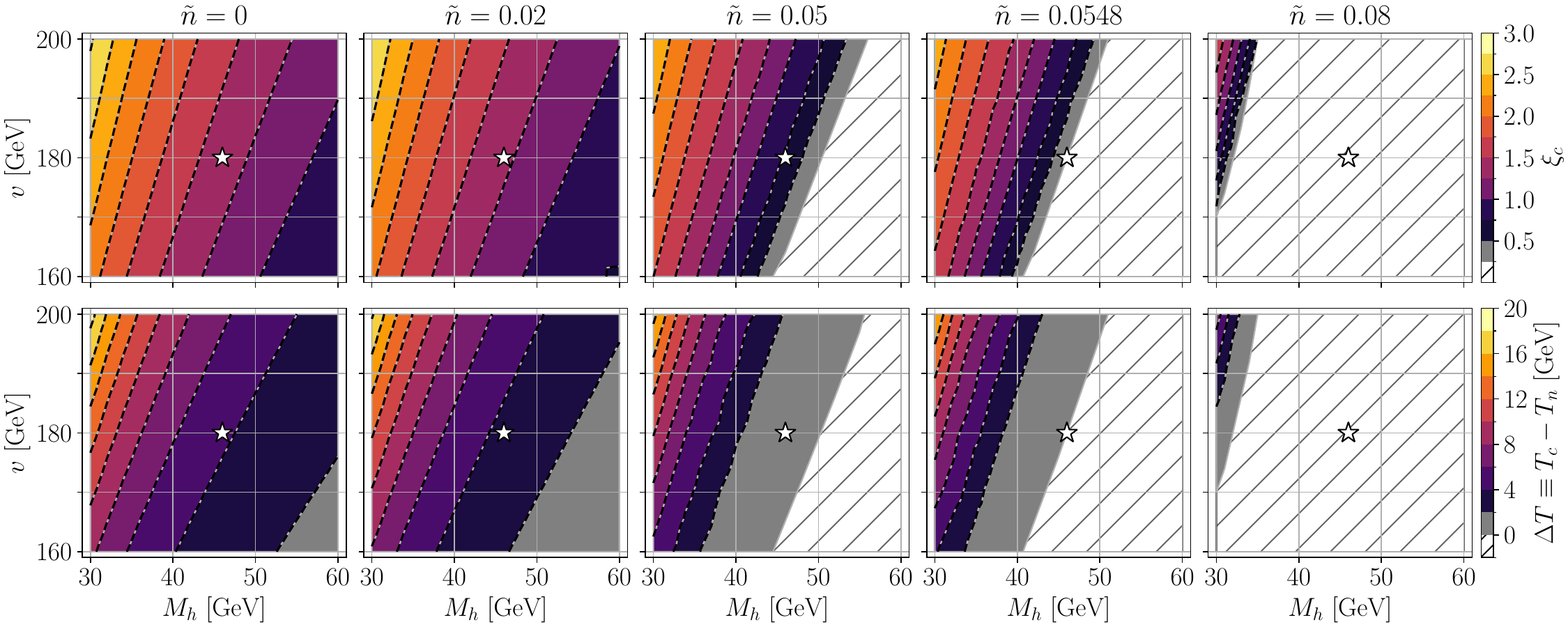}
\caption{Strength of the PT according to Eq.~\eqref{eq:ptstrength} and the difference $\Delta T$ between the critical and the nucleation temperature (second row) in the parameter space defined by $M_h$ and $v$ for values of the PL $\tilde n = \{0,\,0.02,\,0.05,\,0.0548,\,0.08\}$ (columns) and for $\gtwo=\num{1.2}$. The hatched area corresponds to the parameter space regions for which a second-order PT (potentially a smooth cross-over) is found. The BP: $M_h = \SI{46}{GeV}, v = \SI{180}{GeV}, g = \num{1.2}$, given in Eq.~\eqref{eq:bp}, is marked with an asterisk.}
\label{fig:boson_grid_g2_12em1}
\end{figure}

How, then, do the findings for the BP of Eq.~\eqref{eq:bp} translate to the parameter space spanned by $M_h$ and $v$? In Fig.~\ref{fig:boson_grid_g2_12em1}, we illustrate this parameter space for $g=\num{1.2}$ with $\SI{30}{GeV}\leq M_h \leq\SI{60}{GeV}$ and $\SI{160}{GeV}\leq v \leq\SI{200}{GeV}$ and for $\tilde n = \{0,\,0.02,\,0.05,\,0.0548,\,0.08\}$ (columns) with the coloured contours corresponding to the value of $\xi_c$ (top row) and the difference between the critical and nucleation temperature, $\Delta T \equiv T_c - T_n$ (bottom row). We find that our observation for the chosen BP of Eq.~\eqref{eq:bp}, marked with an asterisk in Fig.~\ref{fig:boson_grid_g2_12em1}, directly translates to a larger parameter space. Larger values of $\tilde n$ lead to a decrease in $\xi_c$ that shifts the region of viable (strong) first-order PTs towards even smaller $M_h$ and larger $v$. Consequently, regions with a (strong) first-order PT in the conventional perturbative treatment with $\tilde n = 0$ can show a weak first-order or second-order PT (hatched area) for $\tilde n >0$. In addition, the gap between the critical and nucleation temperatures, as displayed in the bottom row of Fig.~\ref{fig:boson_grid_g2_12em1}, is observed to be reduced for a fixed location in the $(M_h,\,v)$-plane (correlated with the weakened PT), while the total area covered by the $\Delta T$ contours also decreases.\footnote{We find a similar behaviour for the gap between nucleation and percolation temperature.} Consequently, the Polyakov contribution not only drives (strong) first-order PTs towards second-order ones, but also, in line with expectations, reduces supercooling and bubble sizes.

\begin{figure}[!t]
\centering
\includegraphics[width=.645\textwidth]{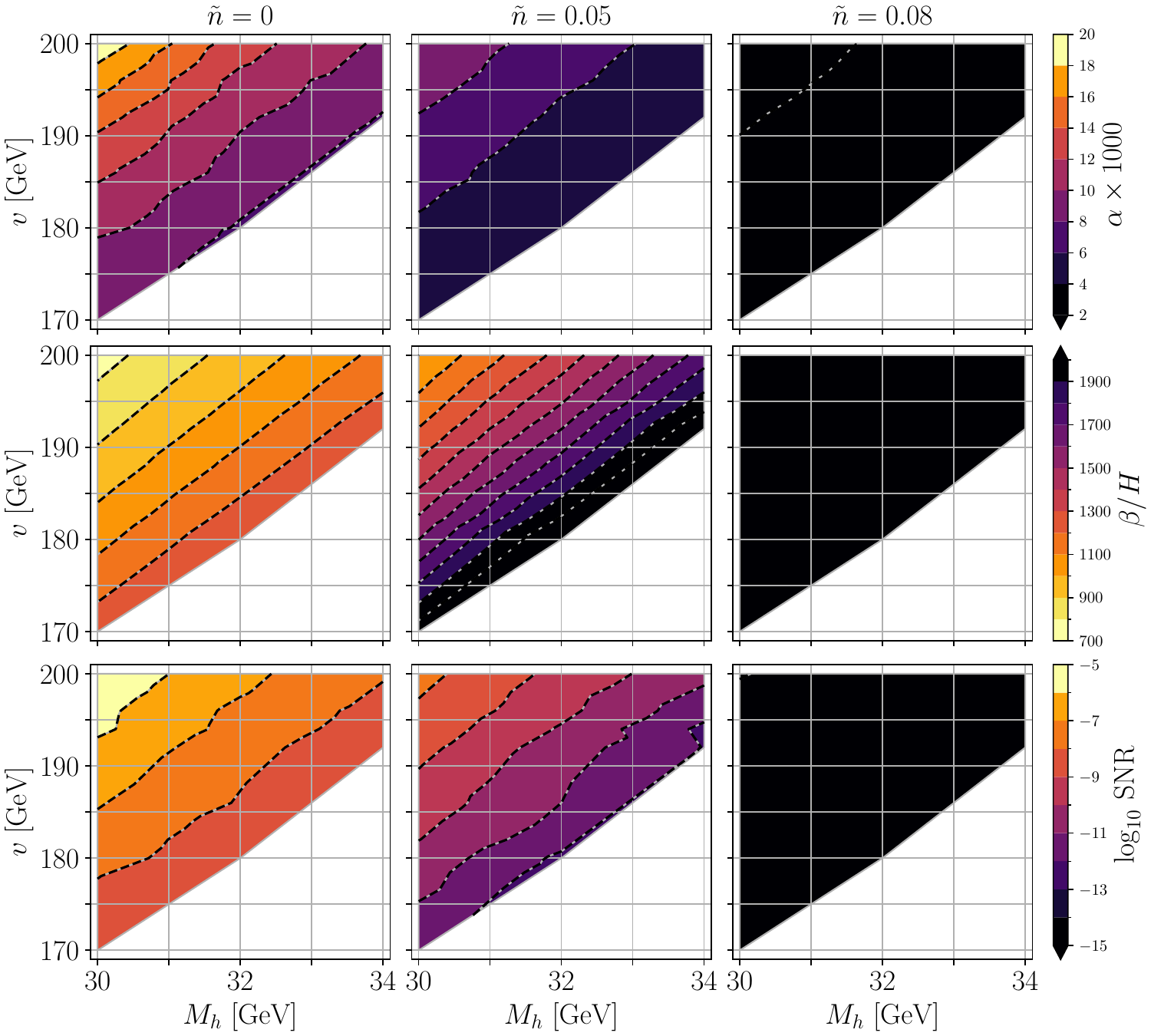}
\vspace*{-1em}
\caption{PT strength $\alpha$, inverse PT duration $\beta/H$ and signal-to-noise ratio (SNR) for detection by LISA (assuming a three-year data-taking period) for $g=\num{1.2}$ and $\tilde n = \{0,\,0.05,\,0.08\}$ in the parameter space region where we still find a first-order PT even for $\tilde n = 0.08$.}
\label{fig:pseudosm_grid_gw}
\end{figure}

As part of a wider programme to track electroweak first-order PTs in the early universe, the relation of such transitions to a stochastic gravitational-wave background is increasingly exploited to connect particle physics with gravitational-wave (GW) astronomy~\cite{Caprini:2019egz,Friedrich:2022cak,Biekotter:2022kgf,Ramsey-Musolf:2024ykk,Biekotter:2025gkp,Conaci:2026djb}.
In the following, we estimate the effect of $\tilde n$ on first-order PT-sourced GW spectra. Using {\tt BSMPTv3.2.1}, we assume GWs to be sourced at the percolation temperature with a wall velocity of $v_w=\num{0.95}$. Thus, the two key parameters for GWs are the PT strength $\alpha$\footnote{This parameter should not be confused with our earlier definition of $\xi_c$; it measures the released vacuum energy density relative to the energy density of a radiation-dominated universe. For the purpose of this study, we estimate $\alpha$ via the trace anomaly~\cite{Hindmarsh:2015qta,Hindmarsh:2017gnf}, but refer the reader to~\cite{Giese:2020rtr,Giese:2020znk} for the more accurate definition in terms of the pseudotrace.} and inverse PT duration $\beta/H$, cf.~Ref.~\cite{Basler:2024aaf} for details on the implementation.
Finally, as an estimate of the detectability of the predicted GWs at the Laser Interferometer Space Antenna (LISA)~\cite{Folkner:1998eni,Danzmann:2000yvf,LISA:2017pwj,LISACosmologyWorkingGroup:2022jok}, we obtain signal-to-noise ratios (SNRs), see Ref.~\cite{Basler:2024aaf} and references therein for details.\footnote{For the chosen toy scenario with $v_w = \num{0.95}$, the dominant source of GWs are sound waves, which we model as a double broken power law~\cite{Caprini:2024hue}, as further detailed in Ref.~\cite{Basler:2024aaf}.}

The results for our toy model are presented in Fig.~\ref{fig:pseudosm_grid_gw}. In line with our observations discussed above, increasing $\tilde n$ decreases the PT strength $\alpha$, and increases the inverse PT duration $\beta/H$, thereby~yielding more rapid PTs. These effects conspire to drastically reduce the SNR, suggesting that Polyakov contributions may be critical for analysing GW signatures of PTs in more realistic models.\footnote{We note that a more realistic choice of $v_w$ is expected to lead to lower SNR values, already at $\tilde n = 0$. However, as our focus here is an illustration of the relative suppression of the SNR due to $\tilde n \neq 0$, we choose an overly optimistic $v_w$ to show that even the strongest possible GWs are drastically damped by $\tilde n \neq 0$.}

\begin{figure}[!t]
\begin{minipage}{1\textwidth}
\begin{minipage}{0.47\textwidth}
  \centering
  \vspace{.95cm}
  \includegraphics[width=0.94\linewidth]{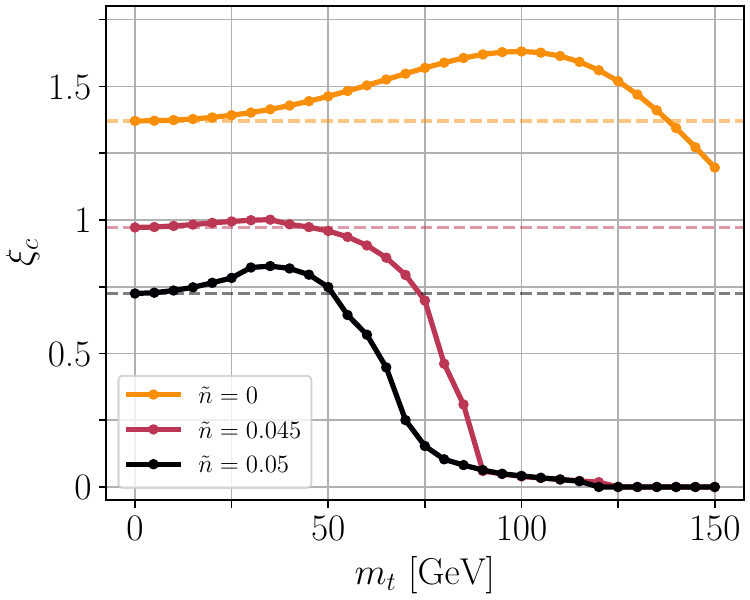}
\end{minipage}
\begin{minipage}{0.52\textwidth}
\begin{small}
    \begin{center}
    \renewcommand{\arraystretch}{1.28} 
    \begin{tabular}{l|ccc}
     \multirow{2}{*}{} & \multicolumn{3}{c}{$\xi_c$} 
     \\
     \cline{2-4}
         $m_t$& $\tilde n = \num{0}$ & $\tilde n = \num{0.02}$ & $\tilde n = \num{0.05}$\\\hline
         \num{0}&\num[round-mode=places,round-precision=3]{1.370483} (\num[round-mode=places,round-precision=3]{0.972228})&\num[round-mode=places,round-precision=3]{1.302365} (\num[round-mode=places,round-precision=3]{0.839684})&\num[round-mode=places,round-precision=3]{0.724468} (0)\\
         \num{5}&\num[round-mode=places,round-precision=3]{1.371257} (\num[round-mode=places,round-precision=3]{0.972581})&\num[round-mode=places,round-precision=3]{1.303088} (\num[round-mode=places,round-precision=3]{0.839935})&\num[round-mode=places,round-precision=3]{0.727417} (0)\\
         \num{25}&\num[round-mode=places,round-precision=3]{1.391787} (\num[round-mode=places,round-precision=3]{0.982805})&\num[round-mode=places,round-precision=3]{1.318641} (\num[round-mode=places,round-precision=3]{0.838939})&\num[round-mode=places,round-precision=3]{0.782812} (0)\\
         \num{50}&\num[round-mode=places,round-precision=3]{1.462719} (\num[round-mode=places,round-precision=3]{1.025583})&\num[round-mode=places,round-precision=3]{1.380391} (\num[round-mode=places,round-precision=3]{0.839346})&\num[round-mode=places,round-precision=3]{0.748479} (0)\\
         \num{75}&\num[round-mode=places,round-precision=3]{1.569037} (\num[round-mode=places,round-precision=3]{1.097760})&\num[round-mode=places,round-precision=3]{1.481874} (\num[round-mode=places,round-precision=3]{0.768462})&\num[round-mode=places,round-precision=3]{0.153477} (0)\\
         \num{100}&\num[round-mode=places,round-precision=3]{1.630878} (\num[round-mode=places,round-precision=3]{1.139872})&\num[round-mode=places,round-precision=3]{1.546404} (\num[round-mode=places,round-precision=3]{0.610061})&\num[round-mode=places,round-precision=3]{0.041569} (0)\\
    \end{tabular}
    \renewcommand{\arraystretch}{1.} 
    \end{center}
\end{small}
\end{minipage}
\caption{Left: $\xi_c$ vs $m_t$ for different choices of $\tilde n$ for the BP with $g=\num{1.2}$ defined in Eq.~\eqref{eq:bp}. Right: Values of $\xi_c$ for the same BP, but for different choices of the gauge coupling: $g = 1.2$ ($g = 1$ in brackets). Second-order PTs, e.g.~for $g = 1$ with $\tilde n = 0.05$, are indicated with $\xi_c = 0$.}
\label{fig:fermion_impact}
\end{minipage}
\end{figure}

\subsubsection*{Fermion-Induced Effects in Phase Transitions}
We conclude our analysis by discussing potential modifications to the critical behaviour outlined in the previous section, arising from the inclusion of chiral fermions that acquire mass via the Higgs mechanism. To this end, we extend the zero-temperature potential by adding
\begin{equation}
V \supset y_t \,\bar Q_L \Phi^c t_R  + y_b\, \bar Q_L \Phi b_R + \text{h.c.}\,,
\end{equation}
with $Q_L = (t_L, b_L)$ and the charge-conjugated field $\Phi^c = i\sigma^2 \Phi^\ast$\,, and
\begin{equation}
y_{t,b} = \frac{\sqrt{2}}{v}m_{t,b}\,,
\end{equation}
where flavour mixing is incorporated through the Cabibbo–Kobayashi–Maskawa (CKM)~\cite{Cabibbo:1963yz,Kobayashi:1973fv}element $V_{tb}$, in analogy with the SM. Additionally, we assume a flavour symmetry among the fermions, to mimic the effect of the SM's colour charge without including QCD gauge boson degrees of freedom.\footnote{Strictly speaking, an additional fermion doublet is necessary to avoid global anomalies~\cite{Witten:1982fp}. If, however, the mass of this fermion is small, it will be phenomenologically irrelevant for the PT as its contribution decouples from the effective potential for a vanishing Yukawa coupling.} This leads to a new contribution to the zero-temperature Coleman-Weinberg potential
\begin{equation}
   V^\mathrm{CW} \supset    \sum_{F= \{t,b\}} \frac{S_F}{64 \pi^2} m_{F}^4  \log\left(\frac{m_{F}^2}{\mu^2} - k_F\right)\,,
\end{equation}
with $S_F \equiv -2 $ and $k_F = 3/2$. The temperature-dependent effective potential receives the additional contributions
\begin{equation}
    V^\beta \supset \sum_{F= \{t,b\}}  
    \dfrac{S_F}{2 \pi^2\beta^4} \, J_{F}(m_X,\beta,\tilde n)\,,
\end{equation}
and the background-field-dependent masses are
\begin{equation}
\label{eq:fermm}
m_{F} = \frac{y_{F}}{\sqrt{2} } \,\omega 
,\quad F\in \{t,b\}\,.
\end{equation}
As there are no fermionic zero modes, no daisy corrections are included.

The combined impact of $\tilde n$ and a non-zero fermion mass on the first-order PT strength $\xi_c$ for our BP of Eq.~\eqref{eq:bp} is shown in Fig.~\ref{fig:fermion_impact}. For any value of $\tilde n$, the inclusion of a non-zero $m_t$ initially yields a stronger first-order PT.\footnote{This aligns with observations in `standard' perturbative studies, cf.~e.g.~\cite{Fok:2008yg,Cao:2021yau}. The inclusion of $m_b\neq 0$ at a fixed $\tilde n$ further amplifies the effect of $m_t\neq 0$. For the inclusion of non-zero mixing, this amplification is weakened. For the purpose of our qualitative study, we limit our following discussion to the case with $m_b = 0$. For a more realistic follow-up, we refer to~\cite{forth}.}
As $m_t$ is increased further, $\xi_c$ first reaches a maximum and then decreases. For sufficiently large $m_t$ and a relatively light scalar mass, the one-loop correction may turn the scalar quartic coupling negative, calling into question the stability of the effective potential. This caps the maximum value of $m_t$ in our analysis. Near this upper limit on $m_t$,\footnote{For the sake of simplicity, we limit our analysis to scenarios where the zero-temperature global NLO minimum coincides with the EW minimum of \SI{246.22}{GeV}. Of course, a sufficiently metastable vacuum also constitutes a physical vacuum.} the first-order PT is found to be weaker than for the purely bosonic case with $m_t = 0$ (for the same $\tilde n$). The increase in first-order PT strength for an intermediate fermion mass range cannot, in general, overcome the weakening effect of the PL. A finite $\tilde n$ is found to push the maximal possible $\xi_c$ to lower $m_t$ and to yield weaker first-order or second-order PTs, for all values of $m_t$.

In addition, we investigate the impact of the gauge coupling $g$. For the relatively small value $g = 1$, as illustrated by the values in brackets in the table on the right side of Fig.~\ref{fig:fermion_impact}, we generally find weaker first-order PTs, compared to the case with $g = \num{1.2}$ (numbers not in brackets), which are further weakened if a finite PL and finite fermion mass are considered. For a choice of $g = 0$, which corresponds to a theory with only scalars and fermions, we generically find very weak first-order, $\xi_c \lesssim \num{.1}$, or second-order PTs.

\section{Summary and Conclusions}
\label{sec:conc}
Electroweak phase transitions in the early Universe provide a phenomenologically rich arena for connecting particle physics with phenomena that remain unexplained by the Standard Model. Given the absence of direct hints of new physics, this research direction has attracted growing attention. This includes work on conceptual problems such as gauge dependence, higher-order corrections, and especially effective-field-theory formulations in thermal field theory \cite{Ekstedt:2024etx,Chala:2024xll,Bhatnagar:2025jhh,Balui:2025yvd,Chala:2025aiz,Chala:2025oul,Chala:2026bar,Bernardo:2026whs,Chakrabortty:2026swu,Bernardo:2026nyq,Bandyopadhyay:2026nrv}. In this work, we study the Polyakov loop~\cite{Polyakov:1975rs,Polyakov:1978vu}, a topologically rooted feature of thermal field theory, and its relation to the effective thermal potential that governs electroweak phase-transition histories. We show that this contribution can have direct implications for associated phenomenological signatures, such as gravitational waves. (For recent analyses in QCD-type setups, yet with similar observations, see also~\cite{Pasechnik:2023hwv,Houtz:2025ogg}.)

We derive the PL term in the effective Higgs potential using the Heat-Kernel method. Treating the PL as a phenomenological parameter, we develop a practical prescription for incorporating these effects into existing perturbative calculations through modifications of the characteristic finite-temperature functions $J_{B/F}$. The standard results are recovered directly in the limit of a vanishing $A_0$ when the PL becomes a trivial identity operator. In this sense, our work provides a first step towards incorporating non-trivial plasma correlations into standard perturbative frameworks~\cite{Wainwright:2011kj,Basler:2018cwe,Athron:2020sbe,Basler:2020nrq,Ekstedt:2022bff,DRansitions,Basler:2024aaf,Athron:2024xrh,Brdar:2025hxw,Costa:2025pew,BLOOP,Matuszak:2026xsz,Wang:2026jjn}.

We apply this approach to a toy model: an $SU(2)$ gauge theory with a complex scalar field in the fundamental representation. This enables us to transparently assess the impact of the PL on the components of the effective potential that control the PT. Starting from regions of the toy-model parameter space that exhibit a first-order PT in the absence of PL contributions, our primary goal is to demonstrate the impact of the PL on the PT dynamics.
Given the close relationship between our scenario and the SM, we expect that many of the implications found in this work will qualitatively extend to the SM. This includes the smoothing of the effective-potential barrier. Overall, phase transitions are tamed by sizeable PL contributions.

We extend our discussion to fermions to clarify the role of heavy quarks as encountered in the SM. To this end, we analyse the intertwined effects of PL contributions and fermion masses in the presence of additional bosonic degrees of freedom. We find that the change in the order and strength of the first-order PT induced by fermionic degrees of freedom follows a pattern closely analogous to the bosonic case, apart from some intermediate oscillatory behaviour.

Throughout our analysis, we consider the PL values as effective phenomenological variables. For a given theory, the PL is fixed by minimising the effective potential with respect to the background scalar and gauge fields. Our analysis suggests that models containing additional non-SM matter fields with non-zero gauge charges will acquire significant contributions from PLs. This can result in the exclusion of large parts of the parameter space that would otherwise be compatible with a (strong) first-order PT. This suggests that such scenarios should be revisited once PL contributions are included consistently. Such effects can alternatively be captured through thermal effective operators where the PL contribution is encoded in the  Wilson coefficients \cite{Chakrabortty:2026swu}. We further expect that higher-order corrections to the thermal effective potential and the inclusion of running parameters in the presence of PL contributions might play important roles in PTs. We leave such a targeted follow-up study to future work~\cite{forth}.

\subsection*{Acknowledgements}
We thank Debmalya Dey, Siddhartha Karmakar, and Philipp Schicho for useful discussions.
We also thank Siddhartha Bandyopadhyay, Goutam Das, Debmalya Dey, Philipp Schicho, and Tushar for valuable comments on the draft. 
LB is supported by the Swiss National Science Foundation (SNSF).
JC acknowledges the hospitality of HRI, Allahabad, India, where part of the research was carried out, and also the support from the Science and Engineering Research Board (SERB), Government of India, under the Project SERB/PHY/2023799.

\appendix
\section{Further details on $\bf\tilde{n}\bf\neq 0$ functions}
\label{sec:moreJ}
The phenomenology that we discuss in this work can be put into an analytical context through an expansion of Eq.~\eqref{eq:stilde} as \cite{Meisinger:2001fi} 
\begin{align}
S^{\tilde{n}}_\Omega(m,\beta)
&= \frac{m^2}{\pi^2\beta^2}\sum_{n=1}^{\infty}\frac{1}{n^2}\cos(2\pi n\tilde{n})\,{\mathbf{K}}_2(mn\beta) \notag\\[4pt]
&= \frac{m^2}{\pi^2\beta^2}\left[\frac{m^2\beta^2}{16}\left\{\ln\!\left(\frac{m\beta\,e^{\gamma_E}}{4\pi}\right) - \frac{3}{4}\right\}\right.  - \frac{1}{2}\left\{\frac{1}{4}\phi^2 - \frac{\pi}{2}\phi + \frac{\pi^2}{6}\right\} \notag\\[4pt]
&\quad + \frac{2}{m^2\beta^2}\left\{-\frac{1}{48}\phi^4 + \frac{\pi}{12}\phi^3 - \frac{\pi^2}{12}\phi^2 + \frac{\pi^4}{90}\right\} \notag\\[4pt]
&\quad + \frac{\pi}{2m^2\beta^2}\sum_{\substack{l\in\mathbb{Z}\\l\neq 0}}\left\{\frac{1}{3}\left[m^2\beta^2 + (\phi - 2\pi l)^2\right]^{3/2}\right. \notag\\[-8pt]
&\qquad\qquad\qquad \qquad  - \frac{1}{3}|\phi - 2\pi l|^3 
 - \frac{1}{2}|\phi - 2\pi l|\,m^2\beta^2
 \left.\left. - \frac{m^4\beta^4}{16\pi|l|}\right\}\right]\,,
\end{align}
where $\phi = [2\pi\tilde{n}]\text{mod}(2\pi)$ with $\tilde{n}<1$.
Furthermore, a closed integral representation can be obtained for $S^{\tilde{n}}_\Omega$ by introducing an imaginary chemical potential, reminiscent of the Popov-Fedotov trick~\cite{Popov:1988fdi} (see also~\cite{Actor:1986zf,Actor:1987cf,Meisinger:2001fi}). Concretely, we can define
\begin{equation}
  \mu(\tilde n) \beta = 2\pi i\,\tilde{n}\,,
\end{equation}
and understand $J_B(m,\beta,\tilde n)$ as the real part of an analytically continued Eq.~\eqref{eq:functionsB0}
\begin{equation}
\label{eq:newJ}
{J}_B(m,\beta,\tilde n) = 
\int_0^{\infty} \d x\, x^2 \,
\text{Re}\left \{ \ln\left(1 - \exp\left\{-\sqrt{x^2 + m^2\beta^2} + \mu(\tilde n) \beta\right\}\right)
\right\}\,.
\end{equation}
$S^{\tilde{n}}_\Omega$ and $J_F$ follow from Eq.~\eqref{eq:newJ} via Eqs.~\eqref{eq:functionsb} and \eqref{eq:functionsFn}.

\bibliographystyle{JHEP}
\bibliography{references}
\end{document}